# Chiral nature of magnetic monopoles in artificial spin ice

N Rougemaille[1], F Montaigne[2], B Canals[1], M Hehn[2], H Riahi[2], D Lacour[2], and J-C Toussaint[1]

[1] Institut Néel, CNRS-UJF, BP 166, 38042 Grenoble Cedex 9, France
[2] Institut Jean Lamour, Université de Lorraine, CNRS, BP 70239, 54506 Vandœuvre lès Nancy, France

E-mail: nicolas.rougemaille@neel.cnrs.fr

**Abstract.** Micromagnetic properties of monopoles in artificial kagome spin ice systems are investigated using numerical simulations. We show that micromagnetics brings additional complexity into the physics of these monopoles that is, by essence, absent in spin models: besides a fractionalized classical magnetic charge, monopoles in the artificial kagome ice are chiral at remanence. Our simulations predict that the chirality of these monopoles can be controlled without altering their charge state. This chirality breaks the vertex symmetry and triggers a directional motion of the monopole under an applied magnetic field. Our results also show that the choice of the geometrical features of the lattice can be used to turn on and off this chirality, thus allowing the investigation of chiral and achiral monopoles.

Modern developments in nanofabrication technology have recently enabled the investigation of fascinating phenomena in frustrated magnetic systems. After the work of Wang and coworkers [1], a wave of studies focusing on artificial realizations of spin ice systems showed that patterned arrays of micro- and nanostructures provide an exciting playground in which the physics of magnetic frustration can be directly observed [2–19]. Through microscopy techniques, this approach offers the appealing opportunity to observe a wide range of phenomena within the concept of lab-on-a-chip and to test theoretical predictions from spin models. Besides the observation of magnetic configurations satisfying the ice rule [12], the evidence of long range dipolar interactions between the magnetic nanostructures [12, 14, 15], or the role of experimental protocols (demagnetization, growth, etc) to approach the ground state manifold [17–19], artificial spin ice systems also allow the study of classical magnetic monopoles [20]. Indeed, if the array is saturated with a magnetic field, it is prepared in a well defined spin ice vacuum. Then, the reversal of a single macro-spin leads to an excess of positive and negative magnetic charges at its two extremities. These two charges can be imaged and split apart under the application of an external magnetic field [21–23], leaving behind them a trace of spin flip events, classical equivalent of Dirac strings [24–29]. These recent results have stimulated new research activities motivated by the quest for magnetic monopoles in condensed matter physics. Even though these monopoles are not the monopoles derived from the Dirac equation [30], they present interesting similarities that open new perspectives in nanomagnetism and magnetic frustration [31].

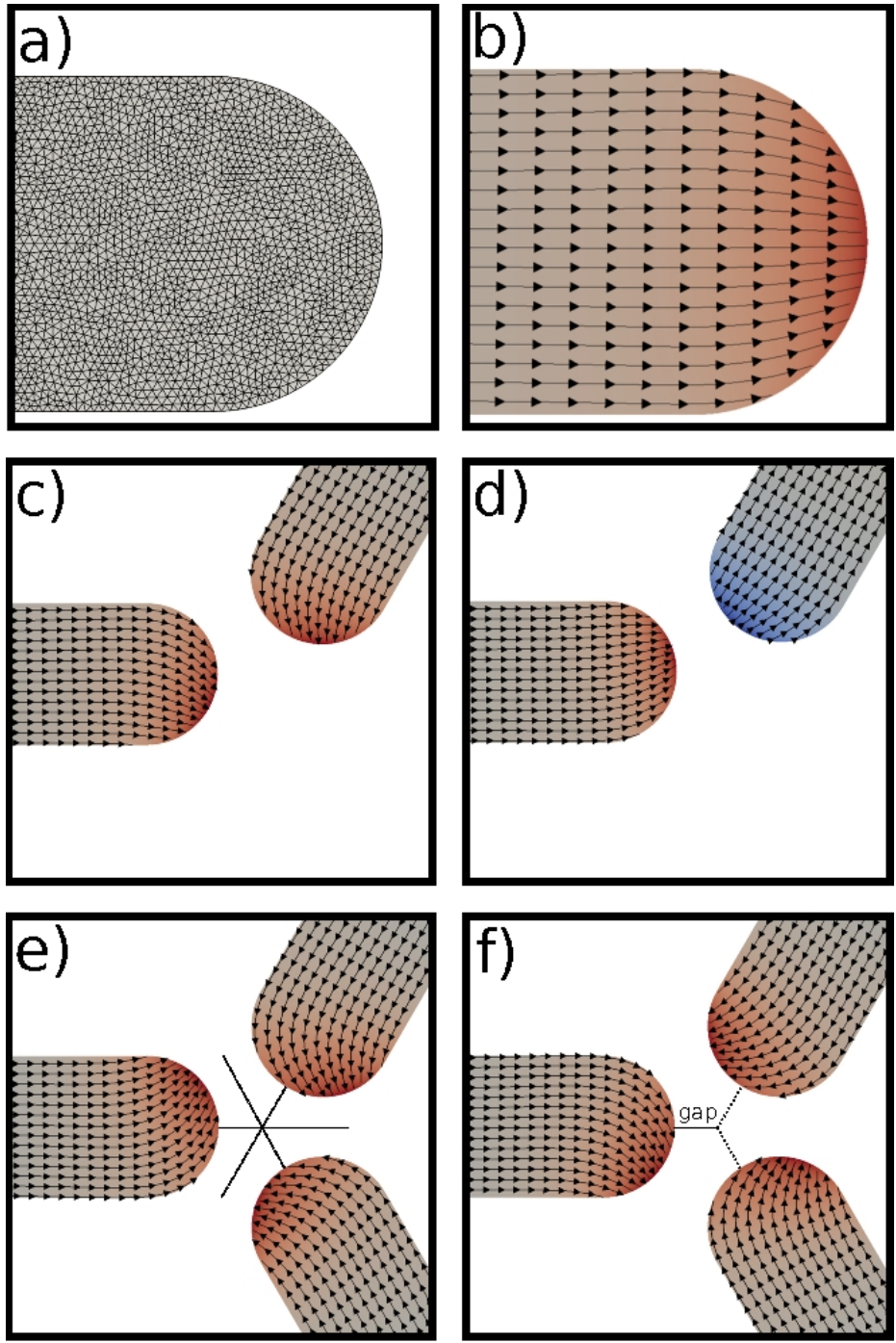

**Figure 1**. (Color online) (a) Illustration of the meshing used in our FE micromagnetic simulations. (b-e) Ground state magnetic configurations at remanence of one (b), two (c-d) and three (e-f) nanomagnets. (b) Result shows that the Ising approximation works well for an individual object. (c-d) For a two element system however, magnetization significantly rotates at the nanomagnets extremities. (e-f) For a three element vertex, forbidden states are chiral with clockwise or anticlockwise chirality. In (c-f) the gap is set to 32 nm in the simulations. Note that the mirror symmetries (indicated by dashed lines) are broken in the three element system due to the chiral nature

of the monopole (e). Color code indicates the amplitude of the magnetization divergence, while the arrows give the local direction of magnetization.

Up to now, in condensed matter systems as well as in artificial arrays of nanomagnets, magnetic monopoles in spin ices have been interpreted as emerging scalar fields, through their relation to the extra magnetic charge carried by a vertex violating the ice rule [20]. In sharp contrast with condensed matter systems, we show in this work that the micromagnetic nature of nanometric pseudo-spins in artificial kagome spin ice introduces the concept of chiral monopoles. A monopole at remanence is then characterized both by a classical magnetic charge and an internal circulation (clockwise or anticlockwise) of a magnetic flux. The chirality can be easily controlled using an applied magnetic field and can be set prior to any motion of the monopole, thus offering an additional degree of freedom to the system. The chirality also breaks the mirror symmetries present in an achiral '3 in' or '3 out' vertex, and gives rise to a directionality of the magnetization reversal process when the monopole is moved under the application of an external magnetic field. Manipulating the chirality of a magnetic monopole thus offer the opportunity to control the trajectory of these emergent classical quasi-particles. Finally, we show that this chirality can be turn on and off by choosing properly the geometrical parameters of the array, allowing investigation of chiral or achiral monopoles.

Our micromagnetic simulations are based on a finite element (FE) approach, i.e. the system is discretized in tetrahedra. This method allows a nearly perfect description of any geometry, and strongly reduces numerical roughness at the edges of the systems, especially on curved shapes (Fig.1a) or in systems that do not have fourfold symmetry. We emphasize that a FE approach is a powerful method for our work as we explore numerically physical effects related to an intrinsic symmetry breaking. We thus want to avoid any artifact induced by numerical roughness associated with finite difference methods, in which the system is discretized into rectangular cells. In this work, we use FEELLGOOD, a homebuilt micromagnetic code [32, 33]. In the following, we consider building blocks of the artificial kagome spin ice made of unconnected cobalt nanomagnets with dimensions 500×100×10 nm$^3$. The gap between the nanomagnets is defined as the distance between the nanomagnet extremities and the center of the vertex (see Fig.1f). The exchange constant is set to 3 pJ/m, the magnetocrystalline anisotropy is neglected, spontaneous magnetization $M_S$ is 1.76 T, mesh size is 4×4×2.5 nm$^3$, and the damping coefficient is set to 1.

In the kagome ice, the number of accessible configurations of a given vertex is $2^3$=8: 6 degenerate low energy configurations ('2 in/1 out' or '2 out/1 in' states), and 2 degenerate high-energy configurations, also called monopoles (because of the magnetic charge in excess with respect to the one of the vacuum), violating the ice-rule ('3 in' or '3 out' states). In artificial spin ice systems, the nanomagnets are not pure Ising spins and the magnetization at their extremity can significantly rotate, especially when the distance between the nanomagnet extremities is small compared to the typical dimensions involved in the system (length or width of the nanomagnets for example). In fact, to minimize the magnetostatic energy at a given vertex, the system always tries to close the magnetic flux. This is illustrated in Figures 1b-f, where the ground state micromagnetic configuration of one, two and three interacting nanomagnets is represented. For a single nanomagnet (Fig.1b), the micromagnetic configuration is symmetric and identical for the 2 extremities: magnetization is essentially uniform but locally follows the curvature of the element: the magnetostatic energy is thus reduced and the exchange energy is increased, but the total energy is lowered. Also, because shape anisotropy is the only source of magnetic anisotropy in our system, magnetization lies in the plane of the nanomagnet, and the out-of-plane component of magnetization is negligible. For two nanomagnets with head-to-head ('2 in') configuration, magnetization distribution is clearly not uniform at the element extremities. As two magnetic charges of same (opposite) polarity give rise to a repulsive (attractive) interaction, magnetization locally rotates in opposite (same) directions (Fig.1c and 1d). For a monopole (Fig.1.e-f), a vortex-like domain wall forms to close the magnetic flux. But two chiralities are possible in that case: the vortex-like configuration can turn clockwise or anticlockwise without changing the system energy, leading to an additional degeneracy of the forbidden states. At remanence, a monopole is then characterized not only by a classical magnetic charge, but also by a chirality.

Interestingly, chiral or achiral monopoles can be prepared by adjusting the gap width. This transition from chiral to achiral monopoles when the gap width is changed simply results from the competition between the magnetostatic interaction, that favors flux closure configurations, and the shape anisotropy that keeps the magnetization aligned along the long axis of the nanomagnets. For large gap widths, the magnetostatic coupling becomes negligible compared to the shape anisotropy, leading to uniformly magnetized nanomagnets and achiral monopoles. On the contrary, small gap widths promote flux closure states. This transition can be visualized by measuring in a given nanomagnet the in-plane component of the magnetization, transverse to its long axis. A non zero value of this component then means that the magnetization is curling at the nanomagnet extremity. On the contrary, a negligible value of this component means that the magnetization is essentially aligned along the long axis of the nanomagnet. In the following, we consider the $<m_y>$ component of the nanomagnet with the long axis aligned along the x direction, as sketched in the inset of Figure 2a, which shows $<m_y>$ computed for different gap distances. $<m_y>$ decreases abruptly with the gap width and is found to be negligible for gap widths larger than 55 nm typically. Above this threshold value (which is peculiar to our geometry and material), monopoles are achiral. Note that such a threshold value is compatible with the resolution of the nanofabrication techniques currently used to elaborate artificial spin ice systems. We can then envision to investigate experimentally the physics of both chiral and achiral monopoles.

We emphasize that the existence of chiral monopoles is a pure micromagnetic effect which is, by essence, absent in spin ice models. This result has one important consequence for artificial spin ice systems: the chirality intrinsically breaks the symmetry of a forbidden vertex. To illustrate this symmetry breaking, we consider in the following one vertex of a honeycomb array where a clockwise monopole has been initialized. We then apply a positive magnetic field of increasing amplitude along the x axis (defined as the long axis of the left hand side nanomagnet, see inset in Fig. 2a) to reverse magnetization in the two other nanomagnets (at 60 degrees from the x axis). We plot the mean value of the $<m_x>$ component of the overall magnetization (i.e. averaged over the three nanomagnets) as a function of the applied field. The striking result is that magnetization reversal is a three step process, as illustrated in Fig. 2b, and one nanomagnet reverses at a much smaller field than the other one. Although the field is applied along one of the three main directions of the vertex, the reversal fields are clearly different for the two (a priori) equivalent nanomagnets. For fields smaller than 35 mT typically, no significant change occurs (bottom left inset): the flux closure state is very stable, and resists the applied field. However, this flux closure state is broken for stronger field (top left inset) and magnetization at the extremity of the nanomagnets then points along the field direction. This switching event corresponds to the first abrupt change observed around 35 mT in Fig.2b. Then, magnetization at the vertex is not characterized anymore by a circulation, and we may think that the memory of the initial chirality is lost when the applied field is larger than 35 mT. A closer look at the micromagnetic configurations reveals however that the magnetic configuration remains asymmetric (as illustrated in the top left inset) and 'remembers' the chirality of the monopole at remanence. For example, magnetization in the nanomagnet with the long axis aligned along the x direction is still pointing upwards (i.e. along y), as in the configuration at remanence. Consequently, reversal is first initiated where magnetization was locally aligned with the applied field (see bottom right inset and corresponding abrupt change around 45 mT in Fig.2b). The vertex then satisfies the ice rule ('2 in/1 out' state) and requires a field stronger than 80 mT typically to reverse the second nanomagnet, leading to another permitted configuration ('2 out/1 in' state corresponding to the third abrupt change in Fig.2b). Here also, note that the field values involved in the two reversal mechanisms are large enough to be accessible experimentally if one wants to evidence anisotropic reversal mechanisms of chiral monopoles.

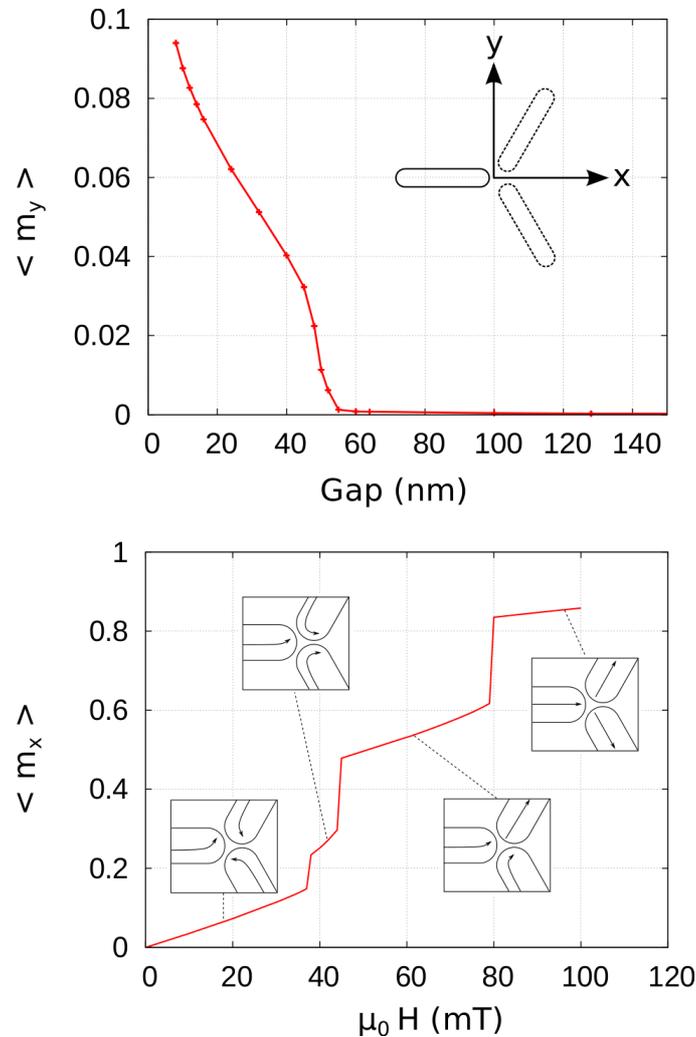

**Figure 2**. (Color online) (a) <My> component of the magnetization as a function of the gap width. (b) <Mx> component of the magnetization as a function of the external field applied along the x direction. Gap is set to 12 nm. See text for details.

This result is important and suggests a possible way to control the chirality of a monopole. The simulations described above show that the micromagnetic configuration of a vertex governs magnetization reversal and is already significantly affected (see top left inset in Fig.2b) by a field smaller than the first switching field. In these simulations, if the applied magnetic field is ramped up to 40 mT and then ramped down to zero, the vertex always restores its initial chirality. As mentioned above, although the chirality is erased by the applied field, the vertex actually keeps the memory of its initial magnetic configuration. If a chiral monopole breaks the system symmetry, the symmetry can also be broken intentionally with a tilted magnetic field. For example, if a 40 mT field is now tilted by a few degrees only with respect to the x axis, the resulting micromagnetic configuration is clearly asymmetric, as illustrated in Figure 3. In particular, the magnetization at the extremity of the nanomagnet aligned along the x axis can be tilted on purpose in the direction of the applied field (see

Fig.3b and 3e compared to Fig.3a and 3d that correspond to the configuration at remanence). In other words, the internal micromagnetic configuration of a monopole can be controlled by an appropriate choice of the applied field. Importantly, the configurations under the applied field are such that two of the three nanomagnets partly close the magnetic flux, thus minimizing locally the magnetostatic interaction. As the field is then ramped down to zero, the local magnetization of the third nanomagnet switches back to stabilize a chiral monopole, as illustrated in Fig.3c and 3f. The key observation here is that the process is deterministic: the local magnetization which reverses is always the one that does not minimize locally the magnetostatic energy (i.e. that does not close the magnetic flux). Depending on which is the third nanomagnet that reverses, the final chirality of the monopole is the same or opposite to the initial chirality [35]. The chirality of a monopole is thus an additional degree of freedom that can be controlled independently of the magnetic charge.

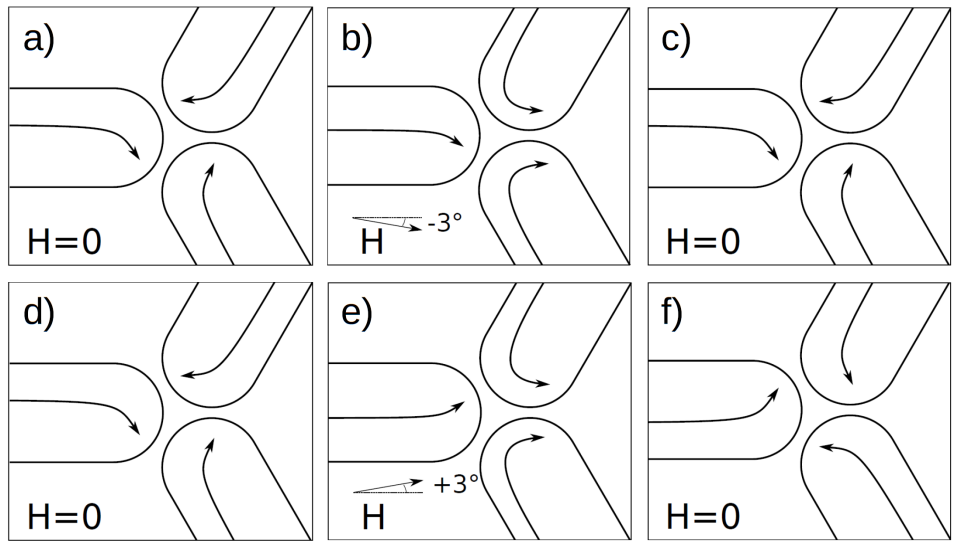

**Figure 3**. Sketches of the internal micromagnetic configuration of a monopole submitted to a magnetic field smaller than the switching field and tilted by ±3 degrees with respect to the x axis. The schematics, deduced from numerical simulations and used to better illustrate the effect, shows that the chirality of a monopole can be controlled before its propagation. See text for details.

One interesting experimental observation in recent literature [21, 22, 28] is that monopoles essentially move along 'straight' lines in the array and do not follow random paths when an external magnetic field is applied. In particular, even when the applied field is aligned along one main direction of the kagome lattice, it was pointed out that magnetization reversal proceeds by a unidimensional anisotropic avalanche, although the two other main directions are expected to be equivalent. We can immediately argue that the applied field is not perfectly aligned with one of the main directions of the kagome lattice and that the experimental uncertainty on the field direction breaks the system symmetry. In the following, we show that the symmetry breaking induced by the chiral nature of the monopoles intrinsically leads to a directionality of the avalanche process. In other words, even in an ideal experiment, with no extrinsic symmetry breaking, monopoles are expected to start their motion in a well-defined direction, determined by the monopole chirality. To do so, we consider a three vertex system made of seven nanomagnets as illustrated in Fig.4.

We then proceed numerically as done experimentally:

- we initialize the system as if it was submitted to a strong magnetic field along the y direction to prepare a saturated state.
- we reverse the direction of one nanomagnet to create a monopole, and we let the system relax. The resulting configuration at remanence is shown in Fig.4a. Note that the monopole has clockwise chirality.
- a 75 mT field is then applied along the -y direction and we let the system evolve.

Consistent with the results shown in Fig.2b, magnetization of the desired nanomagnet reverses, while the rest of the array remains unaffected by the applied field. Magnetization reversal proceeds by the nucleation and propagation of a transverse domain wall (TW) within the nanomagnet [36, 37]. When the TW arrives at the second extremity of the nanomagnet (Fig.4b), the micromagnetic configuration of the second vertex is perturbed by the stray field induced by the TW. Although the external 75 mT field alone is not strong enough to initiate magnetization reversal at the second and third vertices, reversal is initiated when the TW transfers its +2 magnetic charge to the second vertex. Another TW is then nucleated (Fig.4c). Note that the reversal always occurs in the nanomagnet with the long axis aligned along the applied magnetic field (reversal of the other nanomagnet would create a forbidden '3 in' state, which is energetically unfavorable compared to the formation of a configuration satisfying the ice rule). When the second TW arrives at the third vertex, the system has no other choice than transiting via a '3 in' state. The micromagnetic configuration of this monopole is driven by the orientation of the TW and our simulation shows that this monopole induces the next reversal in the same direction of the kagome lattice (Fig.4d) [38]. The process then restarts and an avalanche develops. We emphasize again that there is no intentional symmetry breaking in these simulations and the origin of the directionality of the monopole motion is intrinsic and governed by its chirality at remanence [39].

In sharp contrast with condensed matter systems, we show in this work that monopoles in artificial kagome spin ice systems are not simply scalar quantities characterized by the extra charge carried by vertices violating the ice rule. Instead, micromagnetism in these systems naturally introduces a second quantity, a chirality, which increases the degeneracy of the forbidden states. This chirality can be turned on/off with a proper choice of the lattice geometry and manipulated with an external magnetic field. While at first sight intrinsic monopole propagation could have been expected to follow properties of the 2D random walk, it appears that this new emergent field drives the dimensionality of the monopole propagation within the lattice, leading to a deterministic directionality of the classical Dirac string it leaves behind. We emphasize again that experimentally, extrinsic effects could spontaneously break the system symmetry and likely hide the influence of the chirality on the monopole propagation. This could be the case for example if the applied magnetic field is not perfectly aligned with one of the main directions of the kagome lattice, or if the lattice is slightly distorted across the sample due to imperfections or astigmatism in the nanofabrication process. Whether the intrinsic effect we show in this numerical work can be observed experimentally thus remains an open question. Finally, if we focused our work on the kagome lattice, a similar physics should take place in other geometries. In particular, the "all-in" / "all-out" vertex configurations in the square or triangular geometry should also be characterized by a circulation of the magnetization and break the system symmetry. Although in these two cases monopoles can carry different magnetic charges (+/-2 or +/-4 in the square ice), micromagnetism should also play a key role in magnetization reversal and avalanche mechanisms.

This work was partially supported by the Agence National de la Recherche through projects ANRBLAN2008 'Micro-manip' and ANRBLAN2009 'Spinpress'.

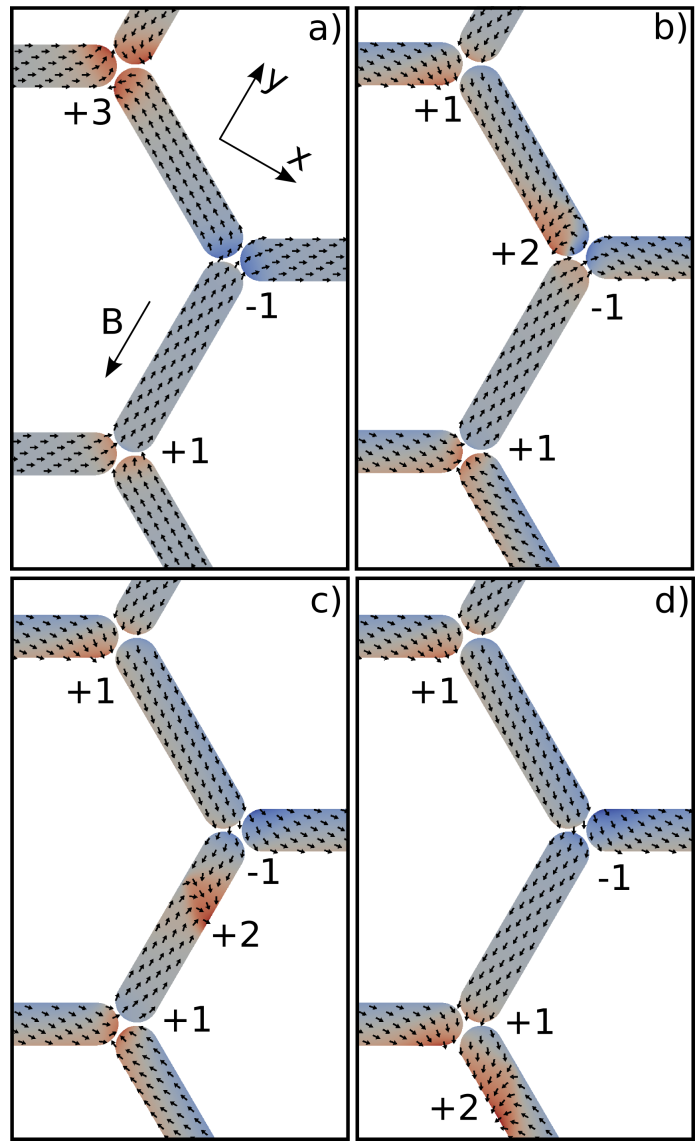

**Figure 4**. (Color online) Snapshots of a micromagnetic simulation showing that a monopole propagates along 'straight' lines of the kagome lattice when submitted to a constant 75 mT magnetic field. The direction of the applied field and the x and y directions are represented by the black arrows in the top left image. The magnetic charge carried by each vertex is also indicated.

III, Topics in Applied Physics, 2006, Volume 101/2006, p.161-205). The dynamics of the TW nucleation, propagation and expulsion of the TW exceeds the scope of this work and will be addressed in a forthcoming publication.